\documentclass[12pt]{article}
\usepackage{amssymb,amsmath,epsfig}

\begin{document}

\title{\bf Effects of the Electromagnetic Field on Five-dimensional Gravitational Collapse}

\author{M. Sharif \thanks{msharif@math.pu.edu.pk} and G. Abbas
\thanks{abbasg91@yahoo.com}\\
Department of Mathematics, University of the Punjab,\\
Quaid-e-Azam Campus, Lahore-54590, Pakistan.}

\date{}
\maketitle

This paper investigates the five-dimensional($5D$) spherically
symmetric gravitational collapse with positive cosmological constant
in the presence of an electromagnetic field. The junction conditions
between the $5D$ non-static interior and the static exterior
spacetimes are derived using the Israel criteria modified by Santos.
We use the energy conditions to discuss solution to the field
equations of the interior spacetime with a charged perfect fluid for
the marginally bound and the non-marginally bound cases. We found
that the range of apparent horizon was larger than that for $4D$
gravitational collapse with an electromagnetic field. This analysis
gives the irreducible and the reducible extensions of $4D$ perfect
fluid collapse with an electromagnetic field and $5D$ perfect fluid
collapse, respectively. Moreover, for the later
case, the results can be recovered under some restrictions.\\


\section*{I. INTRODUCTION}

In general Relativity (GR), the spacetime singularity forms due to
gravitational collapse of a massive astrophysical object. According
to singularity theorems [1], the singularities exist in the form of
either future or past incomplete spacelike geodesics. In other
words, trajectories of material particles (photons) will come to a
sudden end (disappear from the spacetime). The singularity theorems
fail to provide practical information on the nature of the
singularity.

It has been an interesting open problem to know the final fate of
the gravitational collapse. Penrose [2] suggested that the spacetime
singularity is always covered by the event horizon. This is known as
the Cosmic Censorship Hypothesis (CCH), which has no mathematical or
theoretical proof. The spacetime singularity would be a black hole
or a naked singularity, depending on the initial data and the
equation of state. Many efforts have been made to prove or disprove
the CCH. For this purpose, Virbhadra et al. [3] introduced a new
theoretical tool using gravitational lensing. Virbhadra [4] used
gravitational lensing to find an improved form of the CCH.

Oppenheimer and Snyder [5] are pioneers who investigated dust
collapse by taking the static Schwarzschild spacetime as an
exterior spacetime and a Friedmann like solution as an interior
spacetime. They concluded that a black hole is the final fate of
gravitational collapse. Markovic and Shapiro [6] generalized this
work by taking a positive cosmological constant. Lake [7] extended
it for both positive and negative cosmological constants. Sharif
and Ahmad [8] extended spherically-symmetric dust gravitational
collapse with a positive cosmological constant to a perfect fluid.

Recent developments in string theory and other field theories
indicate that gravity is a higher-dimensional interaction. It would
be worthwhile to study gravitational collapse and singularity
formation in higher dimensions. Ghosh and Banerjee [9] have studied
dust collapse with five-dimensional Tolman-Bondi spacetime. In a
recent paper, Sharif and Ahmad [10] studied the gravitational
collapse of a perfect fluid in $5D$ for spherically symmetric
spacetimes.

The behavior of the electromagnetic field in a gravitational field
has been the subject of interest for researchers over the past
decades. The inclusion of an electromagnetic field in gravitational
collapse predicts that the gravitational attraction is
counterbalanced by the Coulomb repulsive force along with the
pressure gradient [11]. Sharma et al. [12] concluded that the
electromagnetic field affects the values of the red-shifts, the
luminosity, and the masses of relativistic compact objects. Nath et
al. [13] demonstrated that the electromagnetic field reduces the
pressure during gravitational collapse. Recently, we [14] found that
the electromagnetic field, along with the matter field, in
gravitational collapse increases the rate of collapse.

It would be interesting to study the gravitational collapse of a
perfect fluid in $5D$ with an electromagnetic field. The junction
conditions between the spherically-symmetric spacetimes are
discussed using the Israel criteria modified by Santos. A closed
form of the exact solution of the field equations in $5D$ exists for
marginally bound $(F(r)=1)$ and non-marginally $(F(r)\neq1)$ bound
cases. This distinguishes the $5D$ case from the $4D$, in which only
a marginally bound solution is possible. The main objectives of this
work are to study the effects of an electromagnetic field on the
rate of collapse in 5$D$ and to see whether or not the CCH is valid
in this framework.

The plan of the paper is as follows: In the next section, the
junction conditions are given. In Section III, we find the $5D$
spherically symmetric perfect fluid solution of the Einstein field
equations with a positive cosmological constant in the presence of
an electromagnetic field for the marginally bound $(F(r)=1)$ and the
non-marginally bound $(F(r)\neq1)$ cases. The apparent horizons and
their physical significance are presented in section IV. We conclude
our discussion in the last section. The geometrized units (i.e., the
gravitational constant $G$=1 and speed of light in vacuum $c=$1 so
that $M\equiv\frac{MG}{c^2}$ and $\kappa\equiv\frac{8\pi
G}{c^4}=8\pi$) are used. All the Latin and Greek indices vary from 0
to 4; otherwise, the indices will be given.

\begin{center}
\section*{II. JUNCTION CONDITIONS}
\end{center}

A timelike $4D$ hypersurface $\Sigma$ is taken such that it
divides a $5D$ spacetime into two $5D$ manifolds, $V^-$ and $V^+$,
respectively. The $5D$ spherically symmetric spacetime is taken as
an interior manifold $V^-$ [15]
\begin{equation}\label{1}
ds_{-}^2=dt^2-X^2dr^2-Y^2(d\theta^2+\sin^2{\theta}d\phi^2
+\sin^2{\theta}\sin^2{\phi}d\psi^2),
\end{equation}
where $X=X(t,r)$ and $Y=Y(r,t)$. For the exterior manifold $V^+$,
we take the $5D$ Reissner-Nordstr$\ddot{o}$m de-Sitter spacetime
\begin{equation}\label{2}
ds_{+}^2=ZdT^2-\frac{1}{Z}dR^2-R^2(d\theta^2+\sin^2{\theta}d\phi^2
+\sin^2{\theta}\sin^2{\phi}d\psi^2),
\end{equation}
where
\begin{equation}\label{3}
Z(R)=1-\frac{M}{R^2}+\frac{Q^2}{2R^3}-\frac{\Lambda}{6}R^2,
\end{equation}
$M$ and $\Lambda$ are constants and $Q$ is the charge. The Israel
junction conditions modified by Santos [16, 17] are
\begin{enumerate}
\item The continuity of the first fundamental form over $\Sigma$ gives
\begin{equation}\label{4}
(ds^2_-)_{\Sigma}=(ds^2_+)_{\Sigma}=ds^2_{\Sigma}.
\end{equation}
\item The continuity of the second fundamental form over $\Sigma$
gives
\begin{equation}\label{5}
[K_{ij}]=K^+_{ij}-K^-_{ij}=0,\quad(i,j=0,2,3,4),
\end{equation}
where $K_{ij}$ is the extrinsic curvature defined as
\end{enumerate}
\begin{equation}\label{6}
K^{\pm}_{ij}=-n^{\pm}_{\sigma}(\frac{{\partial}^2x^{\sigma}_{\pm}}
{{\partial}{\xi}^i{\partial}{\xi}^j}+{\Gamma}^{\sigma}_{{\mu}{\nu}}
\frac{{{\partial}x^{\mu}_{\pm}}{{\partial}x^{\nu}_{\pm}}}
{{\partial}{\xi}^i{\partial}{\xi}^j}).
\end{equation}
$\xi^i$ correspond to the coordinates on ${\Sigma }$ and
$x^{\sigma}_{\pm}$ are the coordinates for $V^{\pm}$. The
Christoffel symbols $\Gamma^{\sigma}_{{\mu}{\nu}}$ are found from
the interior or the exterior spacetime and $n^{\pm}_{\sigma}$ are
components of the outward unit normals to ${\Sigma}$ in the
coordinates $x^{\sigma}_{\pm}$.

The equations of the hypersurface $\Sigma$ in terms of the
coordinates of the interior and the exterior spacetimes are given
as [18]
\begin{eqnarray} \label{8}
h_-(r,t)&=&r-r_{\Sigma}=0,\\
\label{9} h_+(R,T)&=&R-R_{\Sigma}(T)=0,
\end{eqnarray}
where $r_{\Sigma}$ is a constant. Using Eqs. (\ref{8}) and (\ref{9})
in Eqs.(\ref{1}) and (\ref{2}) respectively, it follows that
\begin{eqnarray}\label{10}
(ds_{-}^2)_{\Sigma}&=&dt^2-[Y(r_{\Sigma},t)]^2(d\theta^2+\sin^2{\theta}d\phi^2
+\sin^2{\theta}\sin^2{\phi}d\psi^2), \\\label{11}
(ds_{+}^2)_{\Sigma}&=&[Z(R_{\Sigma})-\frac{1}{Z(R_{\Sigma})}
(\frac{dR_{\Sigma}}{dT})^2]dT^2\nonumber\\
&-&{R_{\Sigma}}^2(d\theta^2+\sin^2{\theta}d\phi^2
+\sin^2{\theta}\sin^2{\phi}d\psi^2).
\end{eqnarray}
For $T$ to be a timelike coordinate, we assume that
\begin{equation}\label{12}
Z(R_\Sigma)-\frac{1}{Z(R_\Sigma)}(\frac{dR_\Sigma}{dT})^2>0.
\end{equation}
From Eqs. (\ref{4}), (\ref{10}) and (\ref{11}), it follows that
\begin{eqnarray}\label{13}
R_\Sigma=Y(r_\Sigma,t),\\
\label{14} [Z(R_\Sigma)-\frac{1}{Z(R_\Sigma)}
(\frac{dR_\Sigma}{dT})^2]^{\frac{1}{2}}dT=dt.
\end{eqnarray}

The outward unit normals to $\Sigma$ in $M^-$ and $M^+$ are
\begin{eqnarray}\label{15}
n^-_\mu&=&(0,X(r_\Sigma,t),0,0,0),\\
\label{16} n^+_\mu&=&(-\dot{R}_\Sigma,\dot{T}, 0,0,0).
\end{eqnarray}
The components of the extrinsic curvature $K^\pm_{ij}$ turn out to
be
\begin{eqnarray}\label{17}
K^-_{00}&=&0,\\
\label{18}
K_{22}^-&=&\csc^2{\theta}K_{33}^-=\csc^2{\theta}\csc^2{\phi}K_{44}^-=(\frac{YY'}{X})_\Sigma,\\
\label{19} K_{00}^+&=&(\dot{R}\ddot{T}-\dot{T}\ddot{R}-\frac{Z}{2}
\frac{dZ}{dR}{\dot{T}}^3+\frac{3}{2Z}\frac{dZ}{dR}\dot{T}{\dot{R}}^2)_\Sigma ,\\
\label{20}
K_{22}^+&=&\csc^2{\theta}K_{33}^+=\csc^2{\theta}\csc^2{\phi}K_{44}^+=(ZR\dot{T})_\Sigma,
\end{eqnarray}
where dot and prime indicate differentiations with respect to $t$
and $r$, respectively. Continuity of the extrinsic curvature gives
\begin{eqnarray}\label{21}
K^+_{00}=0,\quad K^+_{22}=K^-_{22}.
\end{eqnarray}
When we make use of Eqs. (\ref{17})-(\ref{21}) along with Eqs.
(\ref{3}), (\ref{13}) and (\ref{14}), the junction conditions turn
out to be
\begin{eqnarray}\label{22}
(X\dot{{Y'}}-\dot{X}{Y'})_\Sigma=0,\\
\label{23}
 M=(Y^2+\frac{Q^2}{2Y}-\frac{\Lambda}{6}Y^4+{\dot{Y}}^2Y^2
-Y^2\frac{{Y'}^2}{X^2})_\Sigma.
\end{eqnarray}
Equations. (\ref{13}), (\ref{14}) and Eqs. (\ref{22}), (\ref{23})
are the necessary and the sufficient conditions for matching of the
interior and the exterior regions.

\begin{center}
\section*{III. SOLUTION OF THE EINSTEIN FIELD EQUATIONS}
\end{center}

The Einstein field equations with a cosmological constant are
\begin{equation}\label{25}
G^{\mu}_{\nu}-{\Lambda}{\delta}^{\mu}_{\nu}=\kappa(T^{\mu}_{\nu}+E^{\mu}_{\nu}).
\end{equation}
The energy-momentum tensor for a perfect fluid is
\begin{equation}\label{26}
T^{\mu}_{\nu}=({\rho}+p)u^{\mu}u_{\nu}-p\delta^{\mu}_{\nu},
\end{equation}
where $\rho$ is the energy density, $p$ is the pressure, and
$u_\mu$=(1,0,0,0,0) is the 5-vector co-moving velocity.
$E^{\mu}_{\nu}$ is the energy-momentum tensor for the
electromagnetic field given by
\begin{equation} \label{27}
E^{\mu}_{\nu}=\frac{1}{4{\pi}}(-
F^{{\mu}{\omega}}F_{{\nu}{\omega}}+\frac{1}{4}\delta^{\mu}_{\nu}
F_{{\delta}{\omega}}F^{{\delta}{\omega}}).
\end{equation}
First, we solve the Maxwell field equations
\begin{eqnarray}\label{28}
F_{\mu\nu}&=&\phi_{\nu,\mu}-\phi_{\mu,\nu},\\\label{29}
F^{\mu\nu}_{~~;\nu}&=&-4{\pi}J^{\mu},
\end{eqnarray}
where $\phi_{\mu}$ is the five potential and $J^{\mu}$ is the five
current. Because we have taken the charged fluid in a co-moving
coordinate system, the magnetic field will be zero in this case.
Consequently, the five potential and the five current can be taken
as
\begin{eqnarray}\label{30}
\phi_{\mu}&=&({\phi}(t,r),0,0,0,0),\\
\label{31} J^{\mu}&=&{\sigma}\delta^{\mu}_0,
\end{eqnarray}
where $\sigma$ is the charge density.

We treat $\mu$ and $\nu$ as local coordinates for the solution of
Maxwell field equation, Eq. (\ref{29}). Thus, the non-zero
components of the field tensor are evaluated by using Eqs.
(\ref{28}) and (\ref{30}) as
\begin{equation}\label{32}
F_{{t}{r}}=-F_{{r}{t}}=-\frac{{\partial}{\phi}}{{\partial}{r}}.
\end{equation}
Also, Eqs. (\ref{29}) and (\ref{31}) yield
\begin{eqnarray}\label{33}
\frac{1}{X}\frac{{\partial}^2{\phi}}{{\partial}{r}^2}-\frac{{\partial}{\phi}}
{{\partial}{r}}\frac{X'}{X^2}&=&-4{\pi}{\sigma}X,\\\label{34}
\frac{1}{X}\frac{{\partial}^2{\phi}}{{\partial}{r}{\partial}{t}}
-\frac{\dot{X}}{X^2}\frac{{\partial}{\phi}}{{\partial}{r}}&=&0.
\end{eqnarray}
The last equation implies that
\begin{equation}\label{35}
(\frac{1}{X}\frac{{\partial}{\phi}}{{\partial}{r}})=K(r),
\end{equation}
where $K$ is an integrating function and gives the electromagnetic
field contribution. Equations (\ref{33}) and (\ref{35}) yield
\begin{equation} \label{36}
K'(r)=-4{\pi}{\sigma}X.
\end{equation}
The field equations in Eq. (\ref{25}) for the interior spacetime are
written as
\begin{eqnarray}\label{37}
G^{t}_{t}&=&-\frac{3}{X^2}(\frac{Y''}{Y}-\frac{X'}{X}\frac{Y'}{Y}
+\frac{{Y'}^2}{Y^2})+3(\frac{{\dot{Y}}^2}{Y^2}
+\frac{\dot{X}}{X}\frac{\dot{Y}}{Y})+\frac{3}{Y^2}\nonumber\\
&=&\Lambda+K^2+8\pi \rho,\\\label{38}
G^{r}_{r}&=&-\frac{3{Y'}^2}{X^2Y^2}+3(\frac{\ddot{Y}}{Y}
+\frac{{\dot{Y}}^2}{Y^2})+\frac{3}{Y^2}=\Lambda-8\pi
p+K^2,\\\label{39}
G^{\theta}_{\theta}&=&-\frac{2}{X^2}(\frac{Y''}{Y}-\frac{X'}{X}\frac{Y'}{Y}
-\frac{{Y'}^2}{2Y^2})+2(\frac{\ddot{Y}}{Y}+\frac{\dot{X}}{X}\frac{\dot{Y}}{Y}
+\frac{{\dot{Y}}^2}{2Y^2}+\frac{\ddot{X}}{2X})\nonumber\\
&+&\frac{1}{Y^2} =\Lambda-8\pi p-K^2,\\\label{40}
G^{\phi}_{\phi}&=&G^{\psi}_{\psi}=G^{\theta}_{\theta}=\Lambda-8\pi
p-K^2,\\\label{41}
G^{t}_{r}&=&-3\frac{\dot{Y}'}{Y}+3\frac{\dot{X}}{X}\frac{Y'}{Y}=0.
\end{eqnarray}

To solve these equations, we first integrate Eq. (\ref{41}) with
respect to $t$ so that
\begin{equation}\label{42}
X=\frac{Y'}{F},
\end{equation}
where $F=F(r)$ represents the energy inside the hypersurface
$\Sigma$. Making use of Eqs. (\ref{38}) and (\ref{42}), we get
\begin{equation}\label{43}
\frac{\ddot{Y}}{Y}+(\frac{\dot{Y}}{Y})^2+\frac{1-F^2}{Y^2}=\frac{\Lambda-8\pi
p+K^2}{3}.
\end{equation}
The energy-momentum conservation equation for matter,
$T^{\nu}_{\mu;\nu}=0$, shows that the pressure is a function of
$t$ only, i.e., $p=p(t)$. Using this value of $p$ in the above
equation, it follows that
\begin{equation}\label{45}
\frac{\ddot{Y}}{Y}+(\frac{\dot{Y}}{Y})^2+\frac{1-F^2}{Y^2}=\frac{\Lambda-8\pi
p(t)+K^2}{3}.
\end{equation}
Here, we consider $p$ as a polynomial in $t$ given by [8]
\begin{equation}\label{46}
p(t)=p_0(\frac{t}{T})^{- q},
\end{equation}
where $T$ is a constant time introduced in the problem by a
re-scaling of $t$ for physical reasons and $p_0$ and $q$ are
constants. Further, for simplicity, we take $q=0$ so that
$p(t)=p_0$.

Inserting this value of $p(t)$ in Eq. (\ref{45}) and then
integrating with respect to $t$, we have
\begin{equation}\label{46}
\dot{Y}^2=F^2-1+\frac{2m}{Y^2}+(\Lambda-8\pi
p_0+K^2)\frac{Y^2}{6},
\end{equation}
where $m$ is an arbitrary function of $r$ and is related to the
mass of the collapsing system. Using Eqs. (\ref{46}) and
(\ref{42}) in Eq. (\ref{37}), we obtain
\begin{equation}\label{47}
m'=\frac{8\pi}{3}(\rho+p_0)Y^3Y'-\frac{K K'}{6} Y^4.
\end{equation}
For physical reasons, we assume the following energy conditions
[1] are satisfied:
\begin{eqnarray}\label{48}
\rho\geq0,\quad \rho+p\geq0,\\
\rho\geq0,\quad-\rho\leq{p}\leq{\rho}.
\end{eqnarray}
Integrating Eq. (\ref{47}) with respect to $r$, we obtain
\begin{equation}\label{49}
m(r)=\frac{8\pi}{3}\int^{r}_0
(\rho+p_0)Y^3Y'-\frac{1}{6}\int^{r}_0 {K K'}Y^4,
\end{equation}
where $m_0=0$ due to the finite distribution of the mass at the
origin ($r=0$). The mass function $m(r)>0$ because $m(r)<0$ is not
meaningful. Using Eqs. (\ref{42}) and (\ref{46}) in the junction
condition, Eq. (\ref{23}), it follows that
\begin{equation}\label{50}
M=\frac{Q^2}{2Y}+2m+\frac{1}{6}(K^2-{8\pi}p_c)Y^4.
\end{equation}
This mass is located at the origin of the spherical symmetry and
produces a gravitational field in the exterior region of the
sphere.

The total energy $\tilde{M}(r,t)$ up to a radius $r$ at time $t$
inside the hypersurface $\Sigma$ can be evaluated by using the
definition of Misner-Sharp mass [18] given by
\begin{equation}\label{51}
\tilde{M}(r,t)=\frac{1}{2}Y(1+g^{\mu\nu}Y_{,\mu} Y_{,\nu}).
\end{equation}
For the interior metric, it takes the form
\begin{equation}\label{52}
\tilde{M}(r,t)=\frac{1}{2}Y(1+\dot{Y}^2-(\frac{Y'}{X})^2).
\end{equation}
Putting Eqs. (\ref{42}) and (\ref{46}) in Eq. (\ref{52}), we
obtain
\begin{equation}\label{53}
\tilde{M}(r,t)=\frac{m(r)}{Y}+(\Lambda+{K^2}-{8\pi}p_0)\frac{Y^3}{12}.
\end{equation}
Now we solve Eq. (\ref{46}) for the following two cases:\\
$$F(r)=1,\quad F(r)\neq1.$$

\subsection*{1. Solution with $F(r)=1$}

First, we discuss the solution with a positive pressure. For
$\Lambda-8\pi p_0+ K^2>0$, the analytic solutions in closed form can
be obtained from Eqs. (\ref{42}) and (\ref{46}) as follows:
\begin{eqnarray}\label{54}
Y(r,t)&=&(\frac{12m}{\Lambda-8\pi p_0
+K^2})^{\frac{1}{4}}\sinh^{\frac{1}{2}}\alpha(r,t),\\
\label{53}
X&=&(\frac{12m}{\Lambda+{K^2}-{8\pi}p_0})^\frac{1}{4}\left.[\{\frac{4m'}{m}-\frac{2K
K'}{\sqrt{(\Lambda+{K^2}-{8\pi}p_0)}}\}\sinh\alpha(r,t)\right.\nonumber\\
&+&\left.\{{2(t_0(r)-t){KK'}}\sqrt{\frac{3}{2(\Lambda+{K^2}-{8\pi}p_0)}}
+t'_0(r)\sqrt{\frac{(\Lambda+{K^2}-{8\pi}p_0)}{6}}\}\right.\nonumber\\
&\times&\left.\cosh\alpha(r,t)\right.]\sinh^\frac{-1}{2}\alpha(r,t),
\end{eqnarray}
where
\begin{equation}\label{55}
\alpha(r,t)=\sqrt{\frac{2(\Lambda-8\pi p_0+ K^2)}{3}}[t_0(r)-t].
\end{equation}
Here, $t_0(r)$ is an arbitrary function of $r$ and is related to the
time of the formation of the singularity. In the limit
$({8\pi}p_0-K^2) \rightarrow{\Lambda}$, the above solution
corresponds to the $5D$ Tolman-Bondi solution [19]
\begin{eqnarray}\label{56}
\lim_{({8\pi}p_0-K^2) \rightarrow{\Lambda}}Y(r,t)
&=&[8m(t_0-t)^2]^{\frac{1}{4}}, \\\label{57} \lim_{({8\pi}p_0-K^2)
\rightarrow{\Lambda}}X(r,t)
&=&\frac{m'(t_0-t)+2mt_0'}{[32m^3(t_0-t)^2]^\frac{1}{4}}.
\end{eqnarray}

\subsection*{2. Solution with $F(r)\neq1$}

Integrating Eq. (\ref{46}) with the conditions $\Lambda-8\pi
p_0+k^2>0$ and $F(r)\neq1$, it follows that
\begin{eqnarray}\label{58}
Y(r,t)&=&[\{\frac{12m}{\Lambda-8\pi
p_0+K^2}-\frac{9(F^2-1)^2}{(\Lambda-8\pi
p_0+K^2)^2}\}^\frac{1}{2}\sinh\alpha(r,t)\nonumber\\
&-&\frac{3(F^2-1)}{\Lambda-8\pi p_0 +K^2}]^\frac{1}{2}.
\end{eqnarray}
Using Eq. (\ref{58}) in Eq. (\ref{42}), we obtain
\begin{eqnarray} \label{59}
X&=&\frac{1}{2F}[\{\frac{12m}{\Lambda-8\pi
p_0+K^2}-\frac{9(F^2-1)^2}{(\Lambda-8\pi
p_0+K^2)^2}\}^\frac{1}{2}\sinh\alpha(r,t)\nonumber\\
&-&\frac{3(F^2-1)}{\Lambda-8\pi
p_0+K^2}]^{-\frac{1}{2}}[\frac{1}{2}\{\frac{12m}{\Lambda-8\pi
p_0+K^2}-\frac{9(F^2-1)^2}{(\Lambda-8\pi
p_0+K^2)^2}\}^{-\frac{1}{2}}\ \nonumber\\
&\times&\frac{12m'}{\Lambda-8\pi
p_0+K^2}-\frac{24mKK'}{(\Lambda-8\pi
p_0+K^2)^2}-\frac{36FF'(F^2-1)}{(\Lambda-8\pi
p_0+K^2)^2}\nonumber\\
&+&\frac{36KK'(F^2-1)^2}{(\Lambda-8\pi
p_0+K^2)^3}\}\sinh\alpha(r,t)+\{\frac{12m}{\Lambda-8\pi
p_0+K^2}\nonumber\\
&-&\frac{9(F^2-1)^2}{(\Lambda-8\pi
p_0+K^2)^2}\}^\frac{1}{2}\{\frac{2KK'}{6(\Lambda-8\pi
p_0+K^2)^{\frac{1}{2}}}
(t_0(r)-t)\nonumber\\
&+&(\frac{2(\Lambda-8\pi p_0+K^2)}{3})^{\frac{1}{2}}
{t_0}'\}\cosh\alpha(r,t)+\{\frac{6(W^2-1){KK'}}{(\Lambda- 8\pi
p_0+K^2)^2}\nonumber\\
&-&\frac{6FF'}{\Lambda-8\pi p_0+K^2}\}],
\end{eqnarray}
where $\alpha(r,t)$ is given by Eq. (\ref{55}). If the dominant
energy condition holds (i.e., negative pressure), the restriction
$\Lambda-8\pi p_0 + K^2>0$ is removed, and a solution is possible
for all values of $\Lambda$, $p_0$, and $K$. Equations (\ref{58})
and (\ref{59}) represent the non-marginally bound solution
corresponding to $F(r)\neq1$. One can easily verify the marginally
bound solution given by Eqs. (\ref{54}) and (\ref{53}) by
substituting $F(r)=1$ into Eqs. (\ref{58}) and (\ref{59}). The
non-marginally bound solution is impossible in a $4D$
gravitational collapse with an electromagnetic field [14]. For
$K=0$, Eqs. (\ref{54}), (\ref{53}), (\ref{58}), and (\ref{59})
reduce to the marginally bound and the non-marginally bound
solutions of the $5D$ perfect fluid collapse case [10].

\begin{center}
\section*{IV. APPARENT HORIZONS}
\end{center}

In this section, we discuss the apparent horizons and their
physical significance by using the solution to the field
equations. The apparent horizons can be found by using the
boundary of the three trapped spheres whose outward normals are
null. For the interior metric, this is given as follows:
\begin{equation}\label{60}
g^{\mu\nu}Y_{,\mu} Y_{,\nu}=\dot{Y}^2-(\frac{Y'}{X})^2=0.
\end{equation}
Inserting Eqs. (\ref{42}) and (\ref{46}) in the above equation, it
follows that
\begin{equation}\label{61}
(\Lambda+{K^2}-{8\pi}p_c)Y^4-6Y^2+12m=0.
\end{equation}
In particular, when we take $\Lambda=8\pi p_c-{K^2}$,
$Y=\sqrt{2m}$. This is called the Schwarzschild horizon. For
$m=0,~p_0=0$, and ~$K=0$, we have $Y=\sqrt{\frac{6}{\Lambda}}$,
which is called the de-Sitter horizon. The following positive
roots are found from Eq. (\ref{61}).

\textbf{Case (i)}: For $4m<\frac{3}{\Lambda-8\pi p_0 + K^2}$, we
obtain two horizons:
\begin{eqnarray}\label{62}
Y_1&=&\sqrt{\frac{3}{\Lambda-8\pi p_0 +K^2
}+\frac{\sqrt{9-12m(\Lambda-8\pi p_0 + K^2)}}
{\Lambda-8\pi p_0 + K^2}},\\
\label{63} Y_2&=&\sqrt{\frac{3}{\Lambda-8\pi p_0 +
K^2}-\frac{\sqrt{9-12m(\Lambda-8\pi p_0 + K^2)}}{\Lambda-8\pi p_0 +
K^2}}.
\end{eqnarray}
When $m=0$, these reduce to
$Y_1=\sqrt{\frac{6}{(\Lambda+{K^2}-{8\pi}p_c)}}$ and $Y_2=0$.
$Y_1$ and $Y_2$ are called the cosmological horizon and the black
hole horizon, respectively. For $m\neq0$ and
$\Lambda\neq{8\pi}p_c-{K^2},~Y_1$
and $Y_2$ can be generalized [20], respectively. \\\\
\textbf{Case (ii):} For
$4m=\frac{3}{\sqrt{(\Lambda+{K^2}-{8\pi}p_c)}}$, we have repeated
roots:
\begin{equation}\label{64}
Y_1=Y_2=\frac{3}{\sqrt{(\Lambda+{K^2}-{8\pi}p_0)}}=Y,
\end{equation}
which shows that both horizons coincide. The ranges for the
cosmological and the black hole horizons are
\begin{equation}\label{65}
0\leq Y_2\leq{\sqrt \frac{3}{\Lambda-8\pi p_0 +K^2}}\leq Y_1\leq
\sqrt{\frac{6}{\Lambda-8\pi p_0 + K^2}}.
\end{equation}
The black hole horizon has its largest proper area
${4\pi}Y^2=\frac{12\pi}{(\Lambda+{K^2}-{8\pi}p_c)}$, and the
cosmological horizon has its area between
$\frac{24\pi}{(\Lambda+{K^2}-{8\pi}p_0)}$ and
$\frac{12\pi}{(\Lambda+{K^2}-{8\pi}p_0)}$.\\\\
\textbf{Case (iii):} For
$4m>\frac{3}{\sqrt{(\Lambda+{K^2}-{8\pi}p_0)}}$, there are no
positive roots; consequently, there are no apparent horizons.

Now we find the formation time for the apparent horizon with the
help of Eqs. (\ref{54}) and (\ref{61}) given by
\begin{equation}\label{66}
t_n=t_0-\sqrt{\frac{3}{2(\Lambda-8\pi
p_0+K^2)}}\sinh^{-1}(\frac{{Y_n}^2}{2m}-1)^\frac{1}{2},\quad
(n=1,2).
\end{equation}
In the limit $({8\pi}p_0-{K^2})\rightarrow\Lambda$, we obtain the
result corresponding to the $5D$ Tolman-Bondi solution [19],
\begin{equation}\label{67}
t_{ah}=t_0-\sqrt{\frac{m}{2}}.
\end{equation}
Equations (\ref{65}) and (\ref{66}) imply that $Y_{1}\geq Y_{2}$ and
$t_{2}\geq t_{1}$, respectively. The inequality $t_{2}\geq t_{1}$
indicates that the cosmological horizon forms earlier than the black
hole horizon.

\section*{V. OUTLOOK}

This paper provides an extension of our previous analysis for an
electromagnetic field in perfect fluid gravitational collapse with a
cosmological constant [14] from the $4D$ to the $5D$ case. The exact
solution for the interior spacetime with a charged perfect fluid in
the presence of a positive cosmological constant is derived. The
effects of an electromagnetic field on the $5D$ gravitational
collapse are as follows:

The relation for the Newtonian potential is
$\phi=\frac{1}{2}(1-g_{00})$. Using Eqs. (\ref{13}) and
(\ref{50}), for the exterior spacetime, the Newtonian potential
turns out to be 
\begin{equation}
\phi(R)=\frac{m}{{R}^2}+(\Lambda-8\pi p_0 + K^2)\frac{R^2}{12}.
\end{equation}
The corresponding Newtonian force is
\begin{equation}
F=-\frac{2m}{R^3}+(\Lambda-8\pi p_0 + K^2)\frac{R}{6}.
\end{equation}
This force will be zero for

 $R=\frac{1}{(\Lambda-8\pi
p_0+K^2)^\frac{1}{3}}$ and $m=\frac{1}{12(\Lambda-8\pi p_0
+K^2)^\frac{1}{3}}$ while it becomes repulsive (attractive) for
larger (smaller) values of $m$ and $R$. For the repulsive case,
one must have $\Lambda
>({8\pi}p_0-{K^2})$ such that ${8\pi}p_0>{K^2}$ over the entire
range of the collapsing sphere. Notice that $K=K(r)$ gives the
electromagnetic field contribution. From Eq. (\ref{46}), the rate
of collapse turns out be
\begin{equation}\label{86}
\ddot{Y}=-\frac{2m}{Y^3}+(\Lambda-8\pi p_0 + K^2)\frac{Y}{6}.
\end{equation}
Here, we have re-formulated the Newtonian model in terms of the
acceleration of the collapsing process. If $8\pi p_0
-K^2>{\Lambda}$ over the entire range of the collapsing sphere,
then the force becomes attractive, and the cosmological constant
would favor the collapsing process.

Further, we have found two apparent horizons (cosmological and black
hole horizons), whose areas are larger due to the extra dimension as
compared to the values obtained from our $4D$ analysis [14]. The
solution for the non-marginally bound case $(F(r)\neq1)$ is also
possible in $5D$, which is not possible in the $4D$ case. Equation
(\ref{49}) represents the total amount of the collapsing mass
contained in a sphere of radius $0$ to $r$. This amount of mass is
larger than that to the $4D$ case; hence, collapsing process is
faster in this case.

We would like to mention here that all our results match to $5D$
perfect fluid case [10] if we take $K=0$. However, the results for
a $4D$ gravitational collapse with an electromagnetic field [14]
cannot be recovered directly. The reason is that Eq. (61) is
quartic polynomial while in the $4D$ case, the corresponding
equation is a cubic polynomial. In the $5D$ case, the roots under
the possible conditions are different than the roots in $4D$.
Thus, there is only a difference of one degree in both cases, but
the solutions are much different, depending on the nature of the
polynomials.

The cosmological horizon was found to form earlier than the black
hole horizon. Also, Eq. (\ref{66}) shows that the apparent horizon
forms earlier than the singularity; hence, the end state of the
gravitational collapse is a black hole. These evidences agree with
the $4D$ case. There was a possibility of a locally naked
singularity in the $4D$ case due to the presence of an
electromagnetic field, but such a possibility may be avoided due to
the extra dimension. Thus, we conclude that $5D$ favors the
formation of a black hole. It is interesting to mention here that
our study supports the CCH.
\\\\
\vspace{0.25cm}
\begin{center}
{\bf ACKNOWLEDGMENT}
\end{center}
\vspace{0.25cm}
We would like to thank the Higher Education
Commission, Islamabad, Pakistan, for its financial support through
the {\it Indigenous Ph.D. 5000 Fellowship Program Batch-IV}.

\begin{center}
{\bf \large REFERENCES}
\end{center}

\begin{description}

\item{[1]}  S. W. Hawking and G. F. R. Ellis, \emph{The Large Scale Structure of Spacetime}
   (Cambridge University Press, Cambridge, 1979).

\item{[2]} R. Penrose, Riv. Nuovo Cimento \textbf{1}, 252(1969).

\item{[3]} K. S. Virbhadra,  D. Narasimha and  S. M. Chitre, Astron. Astrophys.
\textbf{337}, 1(1998).

\item{[4]} K. S. Virbhadra, Phys. Rev. \textbf{D79}, 083004(2009).

\item{[5]} J. R. Oppenheimer and H. Snyder, Phys. Rev. \textbf{56}, 455(1939).

\item{[6]}  D. Markovic and S. L. Shapiro, Phys. Rev.
\textbf{D61}, 084029(2000).

\item{[7]} K. Lake, Phys. Rev. \textbf{D62}, 027301(2000).

\item{[8]} M. Sharif and Z. Ahmad, Mod. Phys. Lett.
\textbf{A22}, 1493(2007); ibid. 2947.

\item{[9]} S. G. Ghosh and  G. Baneerje, Int. J. Mod. Phys.
\textbf{D12}, 639(2003).

\item{[10]} M. Sharif,  and  Z. Ahmad, J. Korean Phys. Society
\textbf{52}, 980(2008).

\item{[11]} S. Thirukkanesh and  S. D. Maharaj, Math. Meth. Appl. Sci. \textbf{32}, 684(2009).

\item{[12]} R. Sharma,  S. Mukharjee and S. D. Maharaj, Gen. Relativ. Grav. \textbf{33}, 999(2001).

\item{[13]} S. Nath,  U. Debnath, and S. Chakraborty, Astrophys Space Sci. \textbf{313}, 431(2008).

\item{[14]} M. Sharif and G. Abbas, Mod. Phys. Lett. \textbf{A24}, 2551(2009).

\item{[15]} S. G. Ghosh and A. Beesham, Phys. Rev. \textbf{D64}, 124005(2001).

\item{[16]} N. O. Santos, Phys. Lett. \textbf{A106}, 296(1984).

\item{[17]} W. Israel, Nuovo Cimento \textbf{B44}, 1(1966).

\item{[18]} C. W. Misner and D. Sharp, Phys. Rev.
\textbf{136}, b571(1964).

\item{[19]} D. M. Eardley and L. Smarr, Phys. Rev.
\textbf{D19}, 2239(1979).

\item{[20]} S. A. Hayward, T. Shiromizu and K. Nakao, Phys. Rev.
\textbf{D49}, 5080(1994).

\end{description}
\end{document}